\documentclass[fleqn,twoside]{article}
\usepackage{espcrc2}

\usepackage{graphicx}
\usepackage[figuresright]{rotating}

\newcommand{\pinunu}{$K^\pm \rightarrow \pi^\pm \nu \bar{\nu}$}
\newcommand{\AmS}{{\protect\the\textfont2
  A\kern-.1667em\lower.5ex\hbox{M}\kern-.125emS}}

\title{The NA62 RICH detector}

\author{Evelina Marinova\address[MCSD]{INFN - Sezione di Perugia}%
        \thanks{On behalf of the NA62 RICH working group - CERN, Firenze, Perugia, Pisa.}
      }
       
\begin{document}

\begin{abstract}
The RICH detector of the NA62 experiment is proposed for $\pi - \mu$ separation and to contribute to the first level of the trigger. The design parameters of the detector and the results of test beams performed at
CERN in 2007 and 2009 with a prototype are presented.
\vspace{1pc}
\end{abstract}

\maketitle



The main goal of the NA62~\cite{na62} experiment is to measure the branching ratio of the decay \pinunu~ with a precision of 10\% assuming the Standard Model value of the order of $10^{-11}$~\cite{pinunu}.

The main task of the RICH detector is to separate pions from muons in the range 15 -35 GeV/c as the $\mu$-misidentification should be better than $10^{-2}$. In addition to that, it should provide the crossing time of the pion with a resolution better than 100 ps, and an input to Level 0 of the trigger.

The threshold momentum for Cherenkov radiation for pions in Neon (Ne) at atmospheric pressure is $\sim$ 12 GeV/c, which is about 20\% lower than lowest value of the physics range considered. The choice of Ne as radiator guarantees a full efficiency for physics studies. 
The refractive index of the neon at $\lambda = 300$nm is $n = 62.8\times10^{-6}$. The small refractive index of Ne implies a low emission of Cherenkov photons. This is compensated by an 18 m long radiator which is the maximum available space along the beam line.
 Ne has a small dispersion and its effect on the Cherenkov resolution is small.
The small atomic number of Ne is important for reducing the material in front of the electromagnetic calorimeter (LKr). Ne has a good light transparency in the visible and near UV range. 


The RICH detector is composed by a cylindrical vessel which is about 18 m long with a beam pipe passing through it. The vessel volume is $\sim 200~m^3$. Its entrance and exit windows are built of 4 mm thick Al. At the downstream end, there is a mosaic of mirrors pointing to two different regions. The focal length of these mirrors is 17 m. Each of the two focal regions is equipped with 1000 Hamamatsu R7400 U03 PMTs with a pixel size of 18 mm. The two arrays are arranged on the left and on the right side. The density of the gas inside the vessel must remain the same over a long period of time. No circulation of the gas is foreseen, therefore reducing the risk of  temperature changes affecting the gas density (and the refractive index). The pressure variations are expected to be within 0 and 150 mbar overpressure. The RICH operational range starts above 190 nm which makes it very insensitive to impurities. $CO_2$ is used as a transition gas during the filling and the emptying of the vessel with Ne. As the vessel is not able to withstand vacuum, it is first filled with $CO_2$ to replace the air. When the air reaches below the required limits, the gas is circulated in a closed loop and the Ne is filled while absorbing the $CO_2$ in a molecular sieve filter. At the end, the vessel valve is closed.  The level of contamination is expected to be below 1\%. Some tests with Oxygen, Nitrogen and $CO_2$ below 1\% were done to check the performance of the detector.

The mirror mosaic is built of 18 hexagonal and 2 semi-hexagonal mirrors around the beam pipe. The mirrors are made of 2.5 cm thick glass coated with Al. A thin dielectric film is then added in order to improve the reflectivity which should be better than 90 \%. $D_0$ (the diameter of a circle in a focal plane which collects 95\% of the light from a point source) should not be larger than 4 mm. The mirrors support structure has to sustain a total weight of about 400 kg.
 On one hand, a long term stability should be provided. On the other hand, the material should be minimized in order to reduce the interactions before the EM calorimeter. For these reasons, 
a 10 cm thick carbon fiber honeycomb was chosen. Each mirror is supported such that it can be separately adjusted by piezo actuators.

The light detection is done by 2 arrays consisting of 2000 Hamamatsu R7400 U03 PMs. These single anode PMs with metal cover have a peak sensitivity at 420 nm. Their gain is $1.5\times10^6$ at the operating voltage of 900V. They are 18 mm wide while the active area has a diameter of $\sim$ 8mm. Therefore, the light is conveyed by Winston cones~\cite{winston} covered with aluminized mylar foil which is highly reflective. There is a 1 mm thick quartz window separating the air from the neon. To simplify the access to the PMs, they are mounted outside the radiator gas. The PMs are isolated from environmental light by a 1 mm thick O-ring pressed to the quartz window. Each PM HV divider will dissipate 0.3 W which is 300 W for the whole system. A water cooling pipe system will be used to remove the heat . 


The PMT output signal has a roughly triangular shape. A NINO ASIC~\cite{nino} was chosen as time over threshold discriminator due to its intrinsic resolution of 50 ps. To match the optimal NINO performance region, the PMT output is sent to a current amplifier with differential output.
The RICH readout consists of custom made TDC
boards (TDCB)~\cite{tdcb}, equipped with 128 channels of TDC based on HPTDC
chips~\cite{hptdc}. The NINO output signals are sent to FPGA based TELL1 mother
boards~\cite{tell1} housing 4 TDCB (512 channels) each. 
The trigger input will be constructed in parallel with the readout and will be sent to Level 0 of the Trigger. The RICH prototype is the first subsystem available to test the readout system chosen for most of the NA62 detectors.

Two prototypes with the full length of the vessel and a single mirror were tested up to now. The first prototype~\cite{test07} was tested in 2007 and had 96 PMs. Two types of PMs were tested, Hamamatsu R7400 U03/ U06.  PMs  Hamamatsu R7400 U03 were chosen for the final detector due to their better time resolution. The hit multiplicity was measured (17 hits for pion and 6 hits for antiproton), the light collection techniques were tested. The time resolution was measured to be 65 ps  and Cherenkov angle resolution turned out to be about 50 $\mu$rad~\cite{test07}.
\begin{figure}[htb]
\vspace{2pt}
\includegraphics[width=17pc]{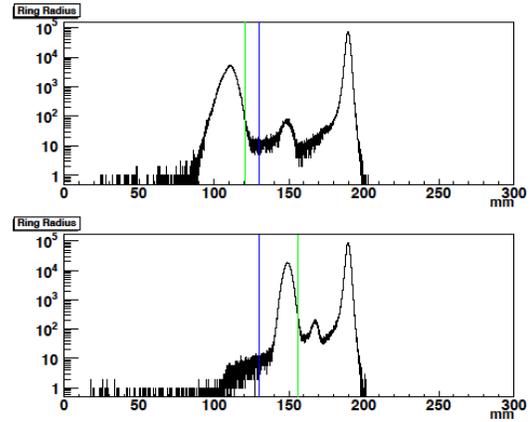}
\caption{Data distribution for the Cherenkov ring radius at 15GeV/c. The green line defines the upper edge of the normalization region. The blue line defines the upper edge of signal region. The leftmost peak is due to $\pi$. The smallest peak is given by true $\mu$ from $\pi$ decays before the beam momentum selection magnets. The rightmost peak is due to positrons. Upper and lower plots are, respectively, for $\pi$ dataset and "$\mu$" dataset. }
\label{fig:pi}
\end{figure}

\begin{figure}[htb]
\vspace{2pt}
\includegraphics[width=17pc]{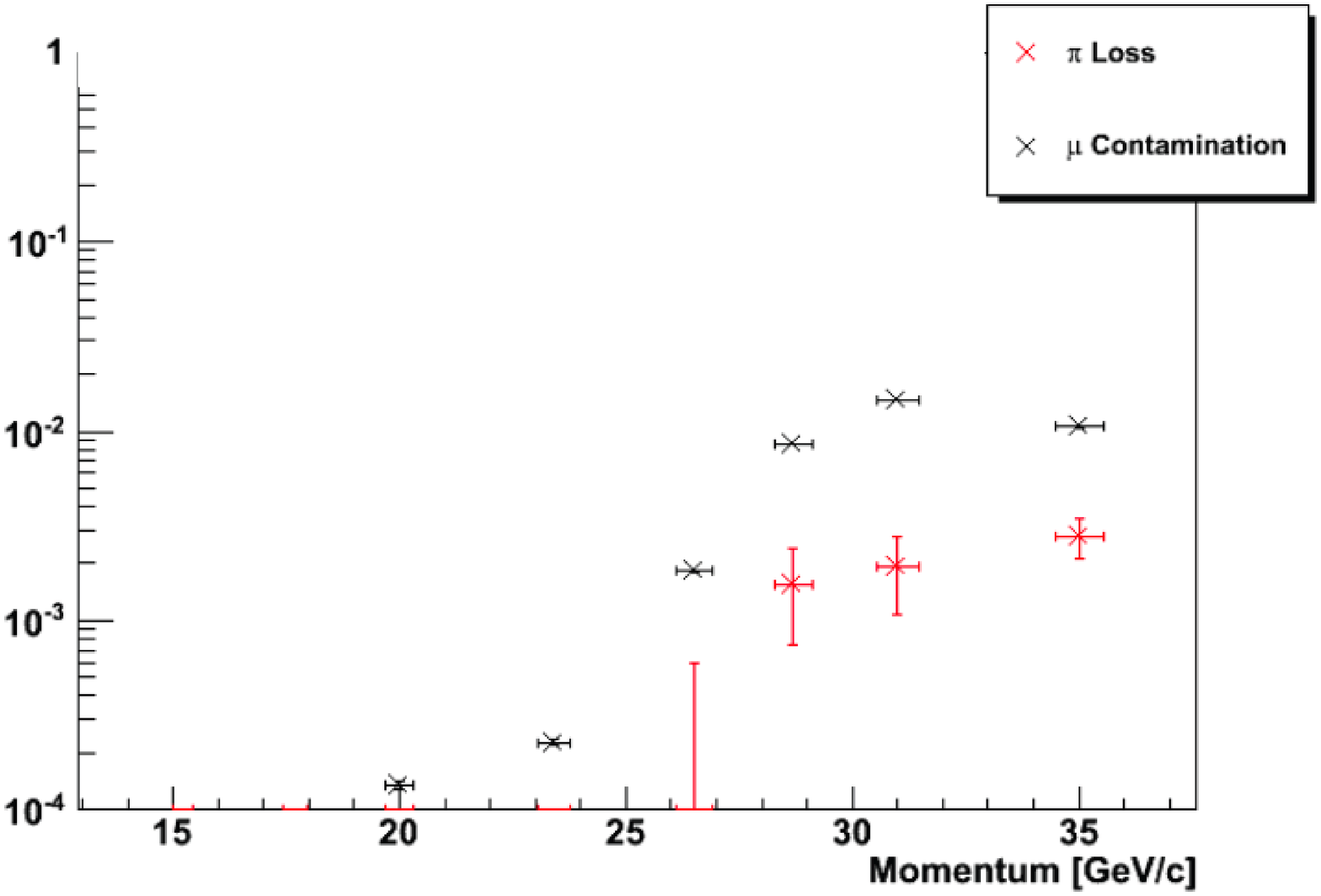}
\caption{Pion loss (red) and $\mu$ mis-identification probability (black) vs the particle momentum.}
\label{fig:prob}
\end{figure}

In 2009, a second prototype with 414 PMs which is 1/5 of the final number was tested~\cite{test09}. The main goals of the test were to check the pion - muon separation, to check the detector performance and to validate the design of the final readout electronics. The prototype performance was tested under different conditions, for example: different mirror positions, different beam rates, different readout firmware versions, different amount of $CO_2$ or air in the Ne. The measurements were repeated by using a new mirror with a better reflectivity which will be used in the final device. Different trigger algorithms and accidental rates were tested at higher intensities.


Positive hadron beams, produced by the SPS primary 400 GeV/c protons was used to produce a secondary positive hadron beam consisting mainly of pions, 15\% protons, few \% K, and variable \% of positrons. 
The beam momentum resolution was $\sim$1\%.The beam momentum was freely selectable in the range 10-75 GeV/c.
 The pion - muon separation was parameterized by the probability to misidentify a muon as a pion.
As it is not easy to have muon beams, the following strategy was adopted.  For each energy measurement, there are two points in the momentum: one at the nominal energy, one for which the pion velocity equals the velocity of the muon.

The distribution of the fitted Cherenkov ring derived from the pions sample at  a given momentum (15.2, 17.7, 20.0, 23.4, 26.5, 28.7, 31.0, 35.0 GeV) was compared to the distributions derived from the "muons"\footnote[1]{ "$\mu$" are $\pi$ at the velocity a $\mu$ would have at the momentum indicated in the $\pi$ set.} sample at the corresponding momentum( 20.0, 23.4, 26.5, 31.0, 35.0, 38.0. 41.0, 46.3 GeV). For example the pions at 20 GeV are considered as "muons" at 15.2 GeV/c . The two samples were used for defining a signal region and to calculate the muon contamination in it. 	

 The Cherenkov ring distribution is shown on Fig.~\ref{fig:pi} where the blue line is half the way between the pion (the left most peak on the upper plot) and the "muon" (the left most peak on the lower plot) peaks. The green line is three standard deviations away from the upper edge of the signal region. The two samples were used for defining a $\pi$ signal region and to calculate the muon contamination in it. The loss of pions is also calculated. The measurements are done for each momentum bin, and then averaged over the momentum range (15 - 35 GeV/c).
The overall integral of these measurements gives a
preliminary muon suppression factor of about 0.7\%(see Fig.~\ref{fig:prob}). 



The time resolution is between 87 ps, at the lower edge of the requested momentum region (15 GeV/c), and 62 ps, at the upper edge (35 GeV/c). The number of hits distribution goes from 8 hits (at 15 GeV/c) to 18 hits (pions at 35 GeV).


In conclusion, the design parameters of the NA62 RICH detector are validated by the results
of the test beams of a full longitudinal scale prototype, carried out at CERN in 2007
and in 2009. 

\end{document}